\documentclass[
reprint,
superscriptaddress,
amsmath,amssymb,
aps,
prb,
]{revtex4-2}
\usepackage[english]{babel}
\usepackage{amsmath}
\usepackage{amssymb}
\usepackage[bookmarks=true,colorlinks,citecolor=blue,linkcolor=blue,allcolors=blue]{hyperref}
\usepackage[capitalise]{cleveref}
\usepackage{physics}
\usepackage[toc,page]{appendix}
\usepackage{graphicx}
\usepackage{orcidlink}

\usepackage{xcolor}

\begin{document}

\title{Gapless Nature of Spin Liquid in the $S=1/2$ Kagome Heisenberg \\ Antiferromagnet with Neural-Network Quantum States}

\author{Andreas Raikos \orcidlink{0009-0006-3203-5445}}
\email{andreas.raikos@irsamc.ups-tlse.fr}
\affiliation{Universit\'e de Toulouse, CNRS, Laboratoire de Physique Th\'eorique, Toulouse,  France.}

\author{Alessandro Sinibaldi \orcidlink{0009-0009-8498-6068}}
\affiliation{Institute of Physics, \'{E}cole Polytechnique F\'{e}d\'{e}rale de Lausanne (EPFL), CH-1015 Lausanne, Switzerland}

\author{Riccardo Rende \orcidlink{0000-0001-5656-4241}}
\affiliation{Center for Computational Quantum Physics, Flatiron Institute, 162 5th Avenue, New York, NY 10010}

\author{Giuseppe Carleo \orcidlink{0000-0002-8887-4356}}
\affiliation{Institute of Physics, \'{E}cole Polytechnique F\'{e}d\'{e}rale de Lausanne (EPFL), CH-1015 Lausanne, Switzerland}

\author{Federico Becca \orcidlink{0000-0002-6142-2050}}
\email{fbecca@units.it}
\affiliation{Dipartimento di Fisica, Universit\`a di Trieste, Strada Costiera 11, I-34151 Trieste, Italy}

\author{Luciano Loris Viteritti \orcidlink{0009-0004-2332-7943}}
\email{luciano.viteritti@epfl.ch}
\affiliation{Institute of Physics, \'{E}cole Polytechnique F\'{e}d\'{e}rale de Lausanne (EPFL), CH-1015 Lausanne, Switzerland}

\date{\today}

\begin{abstract}
    Whether the spin-$1/2$ kagome Heisenberg antiferromagnet hosts a gapped $\mathbb{Z}_2$ or a gapless $U(1)$ Dirac spin liquid remains a central open question in frustrated magnetism.
    Here, we address this question by directly computing the triplet gap with a neural-network quantum state that does not presuppose any specific spin-liquid structure.
 Applying a time-reversal symmetric transformer-based ansatz to clusters of up to $768$ sites, we obtain state-of-the-art variational energies on two-dimensional lattices with periodic boundary conditions and estimate an energy per site of $-0.43873(12)$ in the thermodynamic limit.
    Finite-size scaling of the magnetic order parameter and triplet excitation gap shows that both extrapolate to zero in the thermodynamic limit.
    Together, these results provide independent and direct evidence of a gapless spin-liquid ground state, consistent with the $U(1)$ Dirac spin liquid.
    More broadly, they demonstrate that neural-network quantum states can resolve ground-state properties of frustrated quantum magnets that remain challenging for conventional numerical methods.
\end{abstract}

\maketitle

\paragraph*{Introduction.} 
The $S=1/2$ Heisenberg antiferromagnet on the kagome lattice
\begin{equation}
\hat{H}=J\sum_{\langle i,j\rangle}\hat{\boldsymbol{S}}_i\cdot\hat{\boldsymbol{S}}_j
\label{eq:hamiltonian}
\end{equation}
represents a paradigm for the geometric frustration and a prototypical candidate for hosting a quantum spin liquid ground state~\cite{zheng1990,lecheminant1997,yan2011dmrg}. Here, $J>0$ is the antiferromagnetic super-exchange and $\langle i,j\rangle$ denotes nearest-neighbor pairs on the kagome lattice. The interest in this model was raised following the discovery of several materials featuring localized $S=1/2$ moments that interact via super-exchange within quasi-two-dimensional kagome layers. The most notable example is given by the so-called Herbertsmithite ZnCu$_3$(OH)$_6$Cl$_2$ in which copper ions host $S=1/2$ moments forming kagome layers~\cite{mendels2007,helton2007herbertsmithite,devires2008}. Inelastic neutron scattering has revealed a broad continuum of spin excitations~\cite{han2012fractionalized}, but the nature of its low-energy spectrum remains debated. Neutron magnetic resonance (NMR) and neutron scattering measurements have supported a finite spin gap~\cite{fu2015gapped}, whereas later NMR studies found evidence for gapless excitations~\cite{khuntia2020gapless}. Related kagome compounds have also shown signatures consistent with gapless excitations with conical (Dirac) dispersion~\cite{feng2017barlowite,zeng2024spectral}.

%%%%%%%%%%%%%%%%%%%%%%%%
\begin{figure*}[ht]
\begin{center}\centerline{\includegraphics[width=2.0\columnwidth]{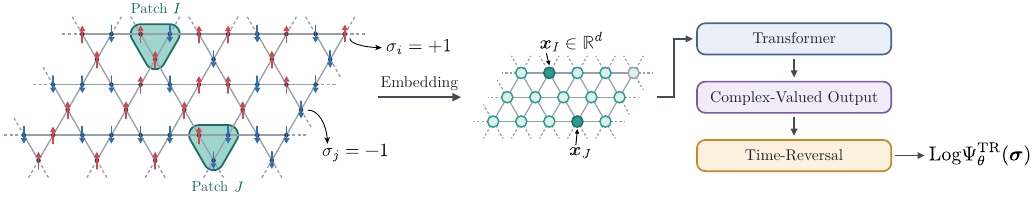}}
        \caption{Schematic of the transformer neural-network wave function used in this work.
        A spin configuration $\boldsymbol{\sigma} = (\sigma_{1}, \ldots, \sigma_{N})$  with $\sigma_i=\pm1$ on the kagome lattice is partitioned into three-site triangular patches, indexed by capital letters.
        Each patch is embedded into a vector $\boldsymbol{x}_I\in\mathbb{R}^d$ on a corresponding triangular lattice. The embedded sequence is processed by a transformer with spatial attention mechanism, followed by a complex-valued output layer and a time-reversal-symmetric layer, yielding the log-amplitude of the variational wave function $\operatorname{Log}\Psi^{\text{TR}}_{\theta}(\boldsymbol{\sigma})$, defined in \cref{eq:time_reversal}.
        }
        \label{fig:artistic}
    \end{center}
\end{figure*}
%%%%%%%%%%%%%%%%%%%%%%%%

Despite the deceptively simple form of the Hamiltonian in~\cref{eq:hamiltonian}, characterizing its ground state remains a formidable challenge. Exact diagonalizations reveal a dense low-energy spectrum~\cite{sindzingre2009gap,lauchli2019exact}, which represents one of the most intriguing aspects of quantum magnets~\cite{mila1998}. Series expansions and entanglement-renormalization calculations favored a valence-bond crystal with a 36-site unit cell~\cite{singh2007vbc,evenbly2010mera}. Subsequent density-matrix renormalization group (DMRG) calculations found evidence in favor of a {\it gapped} spin liquid~\cite{yan2011dmrg,depenbrock2012dmrg}, further characterized by the presence of $\mathbb{Z}_2$ topological order~\cite{jiang2012topological}. Alternatively, variational wave functions, constructed by Gutzwiller-projected Abrikosov fermions, suggested the possibility that the spin liquid is {\it gapless}, with Dirac points in the spinon dispersion and emerging $U(1)$ gauge excitations~\cite{ran2007dirac,iqbal2013kagome,iqbal2014vanishing}. Further support in favor of the existence of gapless excitations was given by tensor-network states, either by pushing the bond dimension to large values (working with projected-entangled pair states)~\cite{liao2017gapless} or by imposing a flux insertion to modify the boundary conditions (within the DMRG approach)~\cite{he2017dmrg}. The dynamical structure factor was also evaluated using DMRG, revealing a low-energy continuum of excitations that is consistent with findings from inelastic neutron scattering~\cite{zhuPNAS}. Other candidate ground states have also been proposed, including a recent claim of a {\it spinon pair density wave}~\cite{duric2025}, which, however, has been challenged due to issues with the underlying Monte Carlo sampling~\cite{kamal2026}. Still, a direct computation of the spin gap (e.g., for $\Delta S=1$ excitations) remained out of reach for large-scale, high-precision numerical methods. Progress in this direction requires controlling both the variational approximation and the approach to the two-dimensional thermodynamic limit. In this respect, neural-network quantum states (NQS)~\cite{carleo2017neural} offer a unique way to reach very high accuracy~\cite{chen2024minsr, rende2024simple} on large clusters~\cite{viteritti2026spatial}, since their representational power can be systematically improved, and the variational state can be optimized directly on two-dimensional lattices with fully periodic boundary conditions. In addition, the implementation of lattice and spin symmetries improves the variational accuracy and enables a direct control of quantum numbers~\cite{nomura2021dirac,nomura2021symmetry,reh2023optimizing,viteritti2025shastry}, so that low-lying excited states can be targeted as the lowest-energy states in different symmetry sectors.

In this Letter, we employ a time-reversal symmetric transformer-based NQS~\cite{viteritti2023transformer,viteritti2025shastry}, optimized within the variational Monte Carlo (VMC) framework~\cite{becca2017quantum}, to study the kagome Heisenberg antiferromagnet of~\cref{eq:hamiltonian} on fully periodic $L\times L$ clusters with $N=3L^2$ spins, up to $N=768$.
For all clusters with $L\geq 6$, we obtain the lowest variational energies reported to date, and the variational accuracy remains approximately constant with system size, as required for a meaningful finite-size scaling.
The magnetic order parameter extrapolates to zero in the thermodynamic limit. Our central result is that the triplet gap, computed directly on large clusters, decreases systematically with system size and extrapolates to zero within error bars in the thermodynamic limit.

Since the transformer ansatz does not presuppose any specific spin-liquid structure, these results provide strong numerical evidence for a gapless spin-liquid ground state consistent with the $U(1)$ Dirac state~\cite{ran2007dirac,iqbal2013kagome,iqbal2014vanishing}.

\paragraph*{Methods.} 
NQS are variational ansätze in which a neural network parametrizes the amplitudes of a quantum many-body wave function, providing a compact representation to approximate the ground state or low-lying excited states in an exponentially large Hilbert space~\cite{rigo2026neural,lange2024architectures,medvidovic2024neural,hermann2023ab}.
To study the Hamiltonian in~\cref{eq:hamiltonian}, we employ an NQS based on the Vision Transformer (ViT) architecture~\cite{viteritti2023transformer,viteritti2025shastry}, augmented with the spatial attention mechanism~\cite{viteritti2026spatial}.
This architecture has been successfully applied to a broad range of strongly correlated spin systems~\cite{viteritti2025shastry,viteritti2026spatial,sinibaldi2026designing,viteritti2025quantum,chen2024minsr,roca2024transformer,raikos2026variational,lu2026kagome3j,rende2026scaling,rende2025}.
In general, the ViT maps a spin configuration $\boldsymbol{\sigma}=(\sigma_1, \dots, \sigma_N)$, with $\sigma_i=\pm1$, to the complex-valued amplitude $\Psi_\theta(\boldsymbol{\sigma})$ in three steps, where $\theta$ denotes the variational parameters.
First, the configuration is partitioned into patches, each of which is embedded into a vector in a $d$-dimensional feature space.
For the kagome lattice, we choose each patch to contain the three spins of one of the $L^2$ kagome primitive unit cells, so that the input to the transformer is the sequence $(\boldsymbol{x}_1, \boldsymbol{x}_2, \ldots, \boldsymbol{x}_{L^2})$ with $\boldsymbol{x}_i \in \mathbb{R}^d$.
Second, this sequence is processed by $n_l$ transformer encoder layers with spatial attention mechanism~\cite{viteritti2026spatial}. Third, the output of the last layer is sum-pooled over the patch index and mapped by a complex-valued nonlinear layer to the amplitude $\Psi_\theta(\boldsymbol{\sigma}) \in \mathbb{C}$.
Further details on each component of the architecture can be found in Ref.~\cite{viteritti2025shastry}.

%%%%%%%%%%%%%%%%%%%%%%%%
\begin{figure}[ht]
    \centering
    \includegraphics[width=\columnwidth]{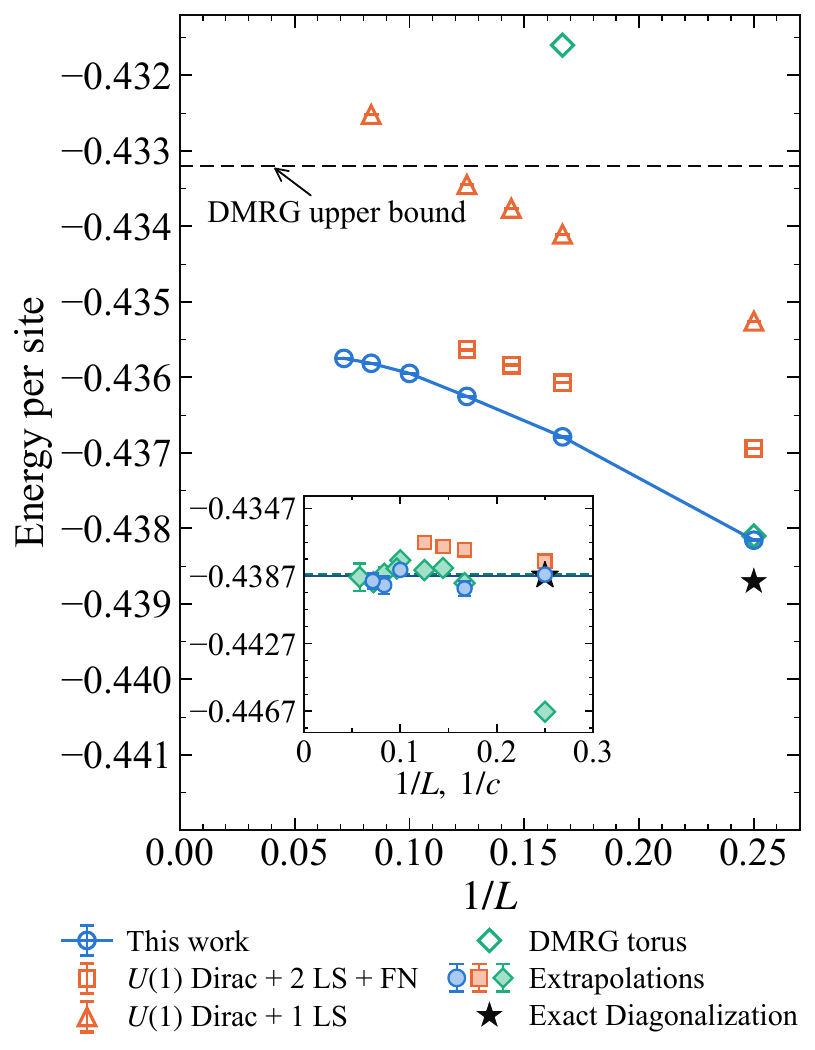}
    \caption{Energy per site of the transformer wave function (blue open circles) as a function of $1/L$, compared with the $U(1)$ Dirac spin liquid improved by one Lanczos step (1~LS, orange open triangles) or by two Lanczos steps plus fixed node (2~LS~+~FN, orange squares)~\cite{iqbal2013kagome,iqbal2014vanishing}, DMRG on $L=4$ and $L=6$ tori (green open diamonds)~\cite{footnote_dmrg,jiang2008dmrg} and exact diagonalization for $L=4$ (black star)~\cite{lauchli2019exact}. Horizontal lines: upper bounds on the thermodynamic-limit energy from DMRG~\cite{yan2011dmrg}.  Inset: zero-variance extrapolations (filled symbols) of our energies, with their constant fit (solid line), and of the $U(1)$ Dirac spin liquid~\cite{iqbal2013kagome} versus $1/L$, and DMRG energies of infinite cylinders [implemented with full $SU(2)$ spin symmetry] versus the inverse circumference $1/c$, with their two-dimensional extrapolation (dashed line)~\cite{depenbrock2012dmrg}.}
    \label{fig:energy}
\end{figure}
%%%%%%%%%%%%%%%%%%%%%%%%

Here, we introduce a simple modification of this architecture that exploits the fact that the Hamiltonian in \cref{eq:hamiltonian} is time-reversal symmetric. Its eigenstates can therefore be chosen real, a property not explicitly imposed by the standard complex-valued ansatz. We enforce this condition directly at the network output by defining the symmetrized wave function
\begin{equation}\label{eq:time_reversal}
    \Psi^{\text{TR}}_{\theta}(\boldsymbol{\sigma}) = \Psi_\theta(\boldsymbol{\sigma}) + \Psi^*_\theta(\boldsymbol{\sigma}) \,,
\end{equation}
which is real by construction and requires neither additional parameters nor additional network evaluations.
This projection both accelerates optimization and improves variational accuracy (see \textit{End Matter}). The complex-valued output layer still encodes the phase structure through the representations learned by the transformer~\cite{viteritti2025shastry}, while phase deviations from $0$ and $\pi$, which a complex-valued ansatz must suppress during optimization~\cite{torlai2020positivization,viteritti2022rbm}, are removed by construction.
The resulting architecture is illustrated in~\cref{fig:artistic}.

The use of patches matching the unit cell, together with translation-equivariant attention and sum pooling over patches, ensures that the ViT amplitude is invariant under Bravais lattice translations of the input spin configuration, with the variational state thus belonging to the zero-momentum ($\Gamma$) sector.
Point-group and spin-inversion symmetries are instead imposed on the network output through quantum-number projection~\cite{viteritti2025shastry,nomura2021dirac}, which improves the variational accuracy and fixes the symmetry sector of the variational state.
For the ground state, we select the even spin-inversion sector ($P=+1$) and the $A_1$ irreducible representation of $C_{6v}$.
For the excited state used to compute the triplet gap (see below), we select the odd sector ($P=-1$) and the $A_2$ representation for $L\geq 6$ ($E_2$ for $L=4$).
Details on the optimization of the ansatz, including the choice of the irreducible representations, are reported in the \emph{Supplemental Material}.

%%%%%%%%%%%%%%%%%%%%%%%%
\begin{figure}
    \centering
    \includegraphics[width=\columnwidth]{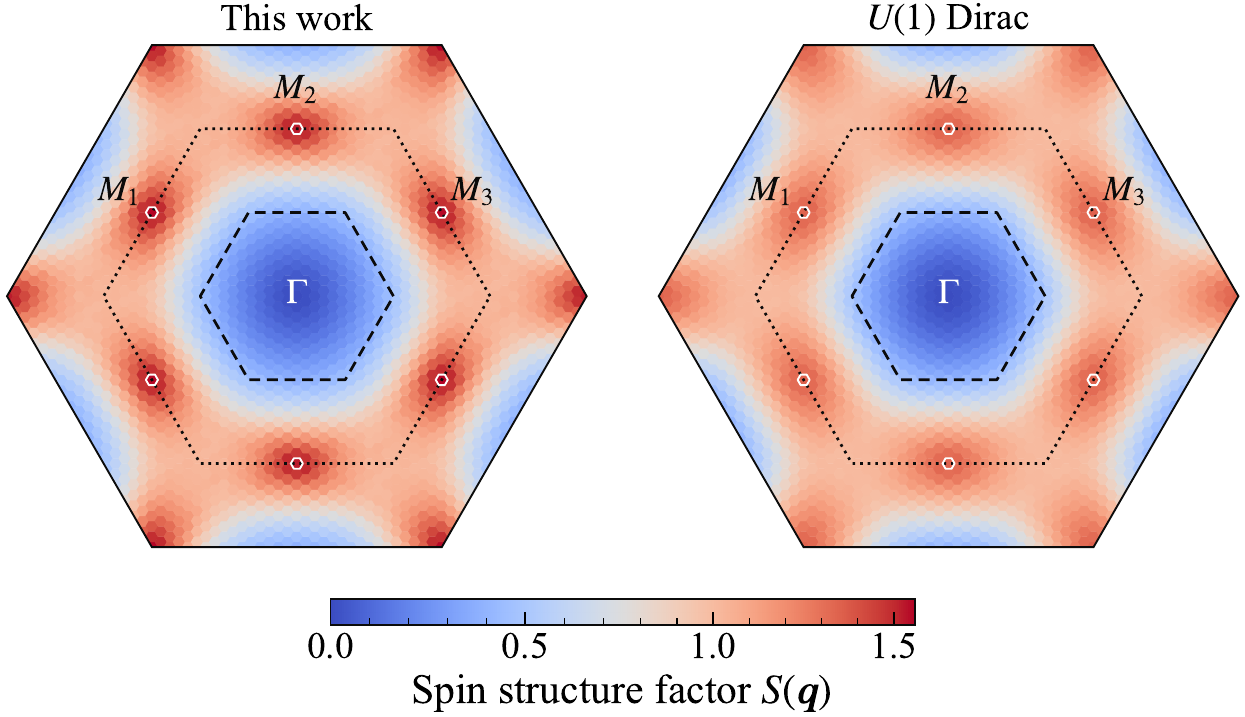}
    \caption{Static spin structure factor $S(\boldsymbol{q})$ (see definition in the main text) of the transformer (left) and $U(1)$ Dirac (right) wave functions on a torus of $N=768$ spins ($L=16$). The dashed (dotted) hexagon represents the first (extended) Brillouin zone. The white markers indicate the six nearest copies of the $\Gamma$ point shifted by reciprocal lattice vectors, which coincide with the $M$ points of the extended zone. Three of them, inequivalent in the extended zone and related by symmetry, are labeled $M_1$, $M_2$, and $M_3$; the other three are their opposites, equivalent to them.}
    \label{fig:sssf_map}
\end{figure}
%%%%%%%%%%%%%%%%%%%%%%%%

\paragraph*{Results.}
We first compare our variational ground-state energies with previous numerical results, restricting the comparison to calculations performed on finite clusters for consistency, see \cref{fig:energy}.
On the smallest cluster, $L=4$ ($N=48$), where exact diagonalizations are available~\cite{lauchli2019exact}, our relative energy error is about $10^{-3}$ (see \textit{End Matter} for additional calculations), comparable to that of DMRG on the same torus with $19\,860$ kept states~\cite{footnote_dmrg}.
Beyond this size, however, DMRG becomes increasingly demanding: maintaining a fixed accuracy in two dimensions generally requires the number of kept states (equivalently, the bond dimension) to grow exponentially with the system width, and fully periodic-boundary conditions impose an additional entanglement cost~\cite{stoudenmire2012dmrg}.
Consequently, on the $L=6$ cluster, DMRG with $6\,000$ kept states~\cite{jiang2008dmrg} already yields a substantially higher energy than ours.
On the larger clusters, our energies are also lower than those of the $U(1)$ Dirac spin liquid improved by one Lanczos step and by two Lanczos steps plus fixed node~\cite{iqbal2013kagome,iqbal2014vanishing}.
\Cref{tab:energy} in the \textit{End Matter} lists all our energies and variances per site, together with the V-score~\cite{wu2024vscore}, a measure of variational accuracy.
The V-score remains approximately constant with increasing $L$, indicating a size-consistent variational accuracy.
Overall, our energies for $L\geq 6$ are, to our knowledge, the lowest variational energies obtained on finite tori. The lower energies reported for the $N=108$ torus in Ref.~\cite{duric2025} may not constitute reliable variational estimates, as a recent work attributes them to nonergodic sampling~\cite{kamal2026}.

We emphasize that, for every size studied, our energies already lie below the upper bound on the infinite-system energy obtained with DMRG~\cite{yan2011dmrg}, shown as horizontal line in~\cref{fig:energy}.
To estimate the ground-state energy in the thermodynamic limit, we perform zero-variance extrapolations at fixed system size (see \textit{End Matter}).
As shown in the inset of~\cref{fig:energy}, the resulting estimates exhibit only a weak size dependence and lie below the corresponding zero-variance estimates for the $U(1)$ Dirac state (improved by Lanczos steps)~\cite{iqbal2013kagome}.
A constant fit of these estimates yields $-0.43873(12)$, compatible within the error bar with the DMRG estimate $-0.4386(5)$ obtained by extrapolating $SU(2)$-symmetric calculations on long cylinders~\cite{depenbrock2012dmrg}.

We next examine the spin-spin correlations in the ground state.
For an exactly $SU(2)$-symmetric singlet, the correlations are identical for all three spin components; since our ansatz fixes only $S^z_{\mathrm{tot}}$ and does not enforce the full $SU(2)$ symmetry, longitudinal and transverse correlations can differ at finite variational accuracy. In the following, we restrict our analysis to the longitudinal ($zz$) correlations rather than the transverse (or isotropic) ones. Indeed, the latter ones exhibit a strong dependence on the accuracy of the wave function (as measured by its energy variance), converging toward the longitudinal values only as the accuracy improves; by contrast, the longitudinal correlations remain largely insensitive when improving the accuracy (see \textit{End Matter}).

%%%%%%%%%%%%%%%%%%%%%%%%
\begin{figure}
    \centering
    \includegraphics[width=\columnwidth]{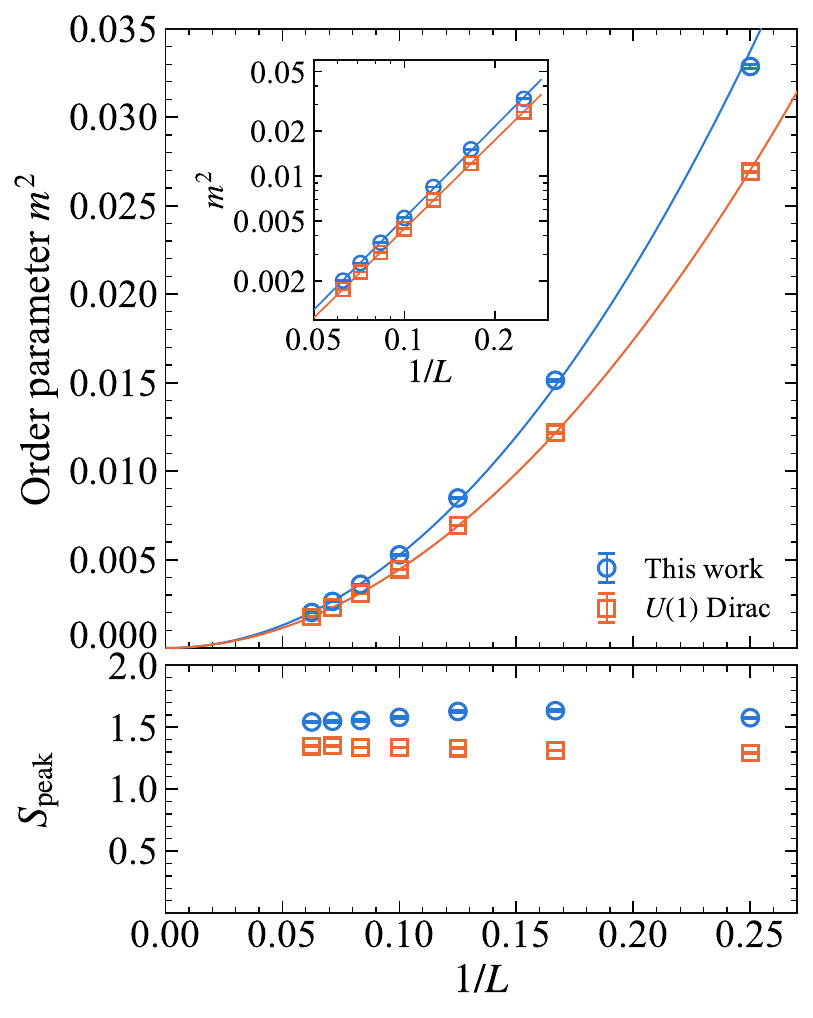}
    \caption{Upper panel: Square magnetic order parameter $m^2$ as a function of $1/L$ for the transformer wave function (blue circles). The solid line is a power-law fit, $m^2\propto L^{-\alpha}$, yielding $\alpha=2.03(2)$. The inset shows the same data on log-log axes. Lower panel: Peak of the spin structure factor $S_{\mathrm{peak}}$ as a function of $1/L$. The definitions of $m^2$ and $S_{\mathrm{peak}}$ are given in the main text. In both panels, the $U(1)$ Dirac state (orange squares) is shown for comparison. Its power-law fit yields a similar exponent close to $2$, and its structure-factor peak shows a similarly weak size dependence.}
    \label{fig:magnetization}
\end{figure}
%%%%%%%%%%%%%%%%%%%%%%%%

%%%%%%%%%%%%%%%%%%%%%%%%
\begin{figure}[t]
    \centering
    \includegraphics[width=\columnwidth]{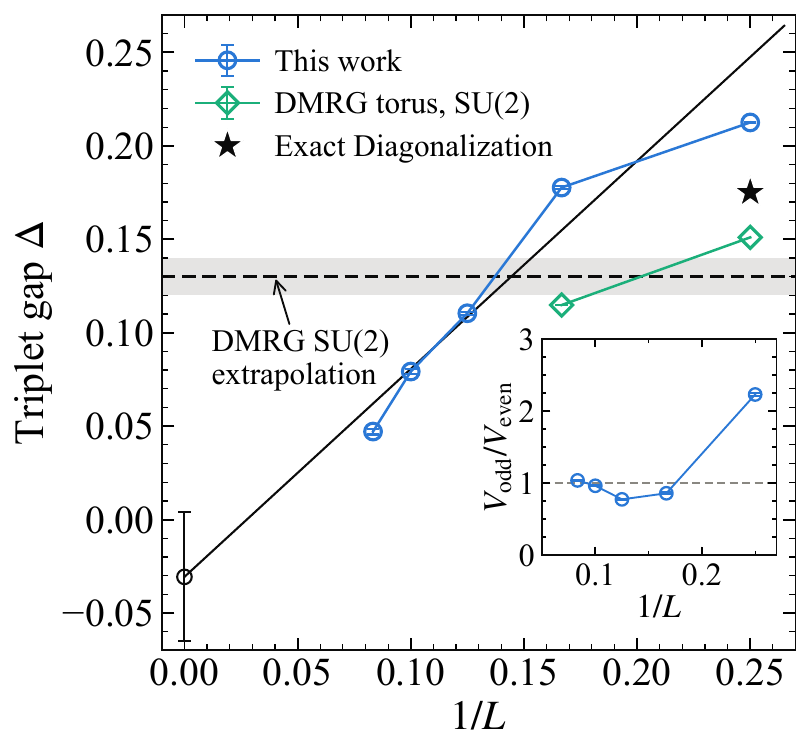}
    \caption{Triplet-gap estimate $\Delta$ as a function of $1/L$ for the transformer wave function (blue circles), compared with DMRG [with full $SU(2)$ spin symmetry] on $L=4$ and $L=6$ tori (green diamonds)~\cite{depenbrock2012dmrg} and exact diagonalization for $L=4$ (black star)~\cite{lauchli2019exact}. The black solid line is a linear fit, extrapolating to $\Delta=-0.031(35)$ in the thermodynamic limit (black circle), consistent with a vanishing gap. The dashed horizontal line and shaded band indicate the thermodynamic-limit estimate $\Delta=0.13(1)$ and its uncertainty, obtained by extrapolating DMRG calculations on long cylinders~\cite{depenbrock2012dmrg}. Inset: ratio of the V-scores~\cite{wu2024vscore} of the odd-parity $V_{\text{odd}}$ ($P=-1$) and even-parity $V_{\text{even}}$ ($P=+1$) states used to compute the gap.}
    \label{fig:gap}
\end{figure}
%%%%%%%%%%%%%%%%%%%%%%%%

We therefore evaluate the spin structure factor as $S(\boldsymbol{q})=\frac{3}{N}\sum_{i,j}e^{\mathrm{i}\boldsymbol{q}\cdot(\boldsymbol{r}_i-\boldsymbol{r}_j)}\langle\hat{S}^z_i \hat{S}^z_j\rangle$.
\cref{fig:sssf_map} shows $S(\boldsymbol{q})$ on the $L=16$ cluster, with maxima at the six symmetry-related $M$ points of the extended Brillouin zone.
To assess whether these peaks signal magnetic long-range order or not, we study the square magnetic order parameter $m^2=S_{\mathrm{peak}}/N$, where $S_{\mathrm{peak}}=\frac{1}{3}[S(M_1) + S(M_2) + S(M_3)]$, with $M_1$, $M_2$, and $M_3$ denoting the three inequivalent, symmetry-related $M$ points (see \cref{fig:sssf_map}).
In \cref{fig:magnetization}, we report $m^2$ as a function of $1/L$. 
The approximately linear behavior of these data on log-log axes (shown in the inset) suggests an algebraic decay of $m^2$ with the system size.
A fit of the form $m^2\propto L^{-\alpha}$ yields an effective exponent $\alpha\approx 2.01(1)$, consistent with $m^2$ vanishing in the thermodynamic limit.
This implies an approximately constant peak of the structure factor $S_{\mathrm{peak}}\propto L^{2-\alpha}\approx \text{const}$.
Remarkably, this scaling closely matches that of the $U(1)$ Dirac spin liquid, shown in the same panel as comparison, indicating that our results are compatible with such a gapless state.

To probe the excitation spectrum more directly, we compute the triplet gap $\Delta$, which is the difference of the total variational energies of the lowest singlet and triplet states.
We target these two states by optimizing wave functions projected onto even ($P=+1$) and odd ($P=-1$) spin-inversion parity, respectively. In both cases, we consider $S^z_{\mathrm{tot}}=0$ and the zero momentum sector, while different point-group symmetries are considered, see \textit{Supplemental Material}. The parity projection separates even and odd total spins: the two states are orthogonal and share no total-spin sector, even though each may still mix several spin sectors of the same parity.

A common alternative is to approximate the lowest singlet and triplet by optimizing at $S_{\mathrm{tot}}^z=0$ and $S_{\mathrm{tot}}^z=1$, respectively~\cite{he2017dmrg,viteritti2025shastry}.
However, fixing the magnetization alone does not exclude triplet components from the $S_{\mathrm{tot}}^z=0$ state and, therefore, the optimization with $S_{\mathrm{tot}}^z=0$ may contain a contribution from the triplet, thus strongly underestimating the triplet gap. This risk is particularly severe on the kagome lattice, where exact diagonalization studies reveal an unusually dense low-energy spectrum in each total-spin sector~\cite{sindzingre2009gap,lauchli2019exact}.

The odd-parity energies are reported in \cref{tab:odd_and_gap} in the \textit{End Matter}, together with their variances and V-scores~\cite{wu2024vscore}.
Because $\Delta$ is the difference of two variational energies, its accuracy depends on how evenly the two states are approximated.
We quantify this balance through the ratio $V_{\mathrm{odd}}/V_{\mathrm{even}}$ of the V-scores of the two states, shown in the inset of \cref{fig:gap}.
Since the V-score is approximately proportional to the relative energy error~\cite{wu2024vscore}, a ratio above one indicates that the triplet is described less accurately than the singlet, so that $\Delta$ is overestimated.
This is the case at $L=4$, where we obtain $\Delta=0.21249(33)$, compared with the exact value $\Delta_{\mathrm{ED}}=0.17505$~\cite{lauchli2019exact}.
There, the V-score ratio is approximately $2.3$, consistent with the ratio of the relative errors with respect to exact diagonalization, and this imbalance accounts for the overestimate (see also \textit{End Matter}).
For $L\geq 6$, the ratio is much closer to one, so we expect the errors of the two energies largely cancel in $\Delta$.

As shown in the main panel of \cref{fig:gap}, the gap decreases systematically with system size, reaching $\Delta=0.0470(16)$ for $L=12$ ($N=432$).
Given the balanced accuracy at these sizes, this decrease is not an artifact of unevenly approximated states.
We also report DMRG results for $L=4, 6$ tori as comparison~\cite{depenbrock2012dmrg}. 
Already at $L=12$, our gap is approximately three times smaller than the thermodynamic-limit estimate $\Delta=0.13(1)$ obtained by extrapolating $SU(2)$-symmetric DMRG calculations on long cylinders~\cite{depenbrock2012dmrg}.
A linear extrapolation in $1/L$ yields $\Delta=-0.031(35)$, consistent with a vanishing gap in the thermodynamic limit.
Together with the scaling of the magnetic order parameter, which closely follows that of the $U(1)$ Dirac spin liquid, our result provides direct evidence for a gapless spin liquid, in agreement with Refs.~\cite{iqbal2013kagome,iqbal2014vanishing,liao2017gapless} and in contrast with the gapped spin liquid proposed in Refs.~\cite{jiang2008dmrg,depenbrock2012dmrg}.

\paragraph*{Conclusions.}
We have studied the spin-$1/2$ Heisenberg antiferromagnet on the kagome lattice using a time-reversal-symmetric transformer-based NQS on fully periodic tori containing up to $N=768$ spins. Our calculations yield state-of-the-art variational energies, while the approximately size-independent V-score~\cite{wu2024vscore} indicates comparable variational accuracy across the system sizes studied. The square magnetic order parameter decreases as a power law, vanishing in the thermodynamic limit with a scaling closely matching that of the $U(1)$ Dirac spin liquid~\cite{iqbal2013kagome}. By optimizing states in distinct symmetry sectors, we also obtain triplet-gap estimates on clusters containing up to $N=432$ spins. The estimated triplet gap decreases systematically with system size and extrapolates to $\Delta=-0.031(35)$ consistent with a vanishing gap. Together, our results provide numerical evidence for a magnetically disordered ground state with gapless excitations, compatible with a $U(1)$ Dirac spin liquid.

Our work demonstrates that NQS can characterize both ground-state correlations and low-energy excitations in challenging two-dimensional frustrated magnets.
Several directions merit further investigation.
Alternative methods for capturing excitations that do not require imposing symmetries, such as grassmamian VMC~\cite{pfau2024accurate,hendry2025grassmannvariationalmontecarlo} and penalty-based approaches~\cite{schatzle2026ab,gao2026excited}, are worth exploring.
Recent advances in fermionic neural-network wave functions~\cite{gu2026solving, roth2025superconductivity, rende2026superconductivity, chen2026extreme} also
motivate extending this approach to the kagome Hubbard model.
More broadly, experimental observations of spin-orbit polarons~\cite{xing2020polaron,chen2024polarons} and of charge order intertwined with superconductivity in kagome metals~\cite{mielke2022charge} make correlated fermionic models on the kagome lattice a compelling target for future studies.
\paragraph*{Code Availability.}
The numerical simulations have been performed using the software libraries NetKet~\cite{netket2,vicentini2022netket} and \texttt{tachys}~\cite{tachys}. A step-by-step tutorial on the implementation of the ViT wave function in \texttt{tachys} is available online~\cite{tachysvit}. The code used to perform the simulations is available at \url{https://github.com/tachys-dev/kagome_vit}.
\paragraph*{Note Added.}
During the preparation of this work, two preprints appeared on arXiv investigating the same model using convolutional neural-network quantum states~\cite{sun2026dirac} and infinite projected entangled-pair states (iPEPS), respectively~\cite{zhang2026}. Both studies report evidence for a gapless spin-liquid ground state, consistent with our conclusions.

\acknowledgments
AR thanks Fabien Alet and Sylvain Capponi for useful discussions. We thank Markus Heyl and Zheng-Hang Sun for comments on the draft. This study has been (partially) supported through the grant NanoX n° ANR-17-EURE-0009 in the framework of the ``Programme des Investissements d'Avenir'' and as part of the ``Swiss AI initiative'' by a grant from the Swiss National Supercomputing Centre (CSCS) under project ID a117 on Alps. We thank CALMIP (Grant No. 2025-P0677 on Olympe cluster, Project No. M25154 on Turpan cluster) and GENCI (Project No. A0190500225) for HPC resources. AS is supported by the Google PhD Fellowship 2025. LLV is supported by SEFRI under Grant No. MB22.00051 (NEQS - Neural Quantum Simulation).
The Flatiron Institute is a division of the Simons Foundation.

\bibliography{refs}

\clearpage

\section*{End Matter}
\suppressfloats[t] % keep the End Matter heading at the top of the page
\paragraph*{Time Reversal Symmetry.}
The Heisenberg Hamiltonian in \cref{eq:hamiltonian} is time reversal symmetric, which allows us to choose its eigenstates to be real. We exploit this property by defining the output of the neural network in the projected form in \cref{eq:time_reversal}.
In addition, before taking the real part in the time-reversal symmetrization, we multiply the ViT output by a fixed phase factor~\cite{capriotti1999triangular}, $\Psi_\theta(\boldsymbol{\sigma})\rightarrow e^{i\Phi_0(\boldsymbol{\sigma})}\Psi_\theta(\boldsymbol{\sigma})$. 
Here $\Phi_0(\boldsymbol{\sigma})=-\frac{1}{2}\sum_i\phi_i\sigma_i$ is the phase of the classical coplanar $\boldsymbol{q}=\boldsymbol{0}$ state and $\phi_i\in\{0,2\pi/3,-2\pi/3\}$ is the angle of the spin at site $i$, which is the same on all sites of a given sublattice.
This phase factor is introduced to accelerate convergence and does not restrict the variational state, since the network can freely modify the sign structure through the phase of $\Psi_\theta(\boldsymbol{\sigma})$.
We have verified that optimizations without it eventually converge to the same results.

Using real amplitudes offers two main advantages. First, the logarithmic derivatives with respect to the variational parameters are also real. This halves the dimension of the sample-space linear system solved by Stochastic Reconfiguration (SR)~\cite{sorella1998green, sorella2005wave}-based methods to compute the parameter update~\cite{chen2024minsr,rende2024simple}, yielding an empirical three- to four speedup per optimization step.

\begin{figure}[b]
    \centering
    \includegraphics[width=\columnwidth]{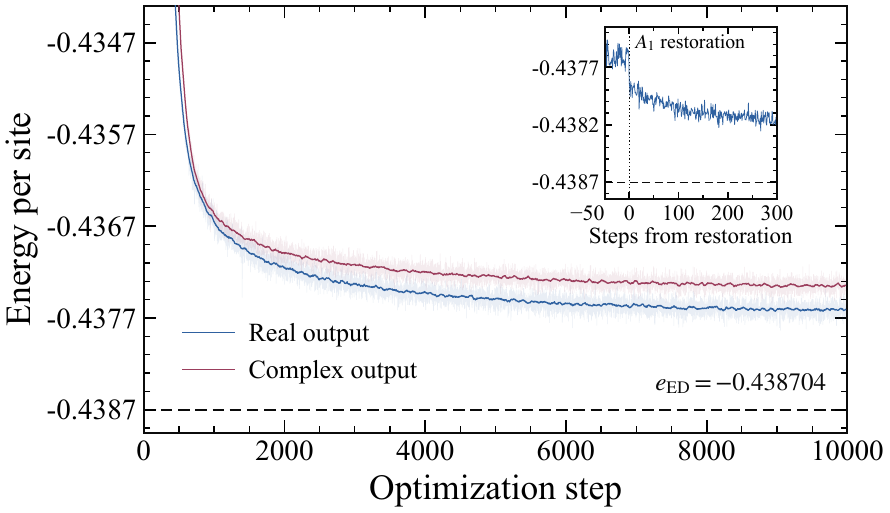}
    \caption{Energy per site during the optimization of real- and complex-output transformer with $n_l=8$ layers, $h=12$ heads and $d=72$ embedding dimension on the $L=4$ cluster ($N=48$). Darker lines show moving averages. The real-output optimization is approximately $\times3$ faster per step than the complex-output optimization. Dashed horizontal lines mark the exact ground-state energy~\cite{lauchli2019exact}. Inset: the final 50 steps of the real-output optimization and its continuation after projection onto the fully symmetric $A_1$ point-group sector. The dotted vertical line marks the onset of symmetry restoration at step zero.
}
    \label{fig:optimization}
\end{figure}

Second, it improves the variational accuracy achieved in practice. To illustrate this, \cref{fig:optimization} compares the optimization of the same architecture with and without the time-reversal-symmetric output layer defined in \cref{eq:time_reversal}, on the $L=4$ cluster ($N=48$). The real-output network converges to a lower variational energy while requiring approximately one-third of the time per optimization step.

%%%%%%%%%%%%%%%%%%%%%%%
%%%%%%%%%%%%%%%%%%%%%%%

\paragraph*{Numerical Values of Energies and Variances.}
\cref{tab:energy} and ~\cref{tab:odd_and_gap} report the final energy per site, variance per site, and V-score after point-group restoration for states with even (up to $L=14$) and odd spin-inversion parity, respectively. At $L=16$, we report the values before the point-group projection, since the cost of becomes considerable at this size.  The corresponding V-scores are denoted by $V_{\text{even}}$ and $V_{\text{odd}}$. All quantities are estimated from approximately $10^6$ Monte Carlo samples per state, generated using blurred sampling~\cite{wan2026blurred} to obtain an unbiased estimate of the energy variance.\\

%%%%%%%%%%%%%%%%%%%%%%%%%%%%%%
\begin{table}[]
    \caption{Variational energy per site, variance per site and V-score ($V_{\text{even}}$) of the ViT states optimized in the ($\Gamma, S^z=0, P=+1$) sector, followed by $300$ optimization steps in the $A_1$ irrep of the $C_{6v}$ point group. For $L=16$, marked by $(*)$, this second stage is omitted
due to computational cost.}
    \label{tab:energy}
    \begin{ruledtabular}
    \begin{tabular}{cccc}
$L$ & Energy per site & Variance per site & $V_{\text{even}}$ \\
\hline
4 & $-0.438156(4)$ & $0.00139(1)$ & $0.00722(4)$ \\
6 & $-0.436788(4)$ & $0.00268(4)$ & $0.0141(2)$ \\
8 & $-0.436252(3)$ & $0.00404(2)$ & $0.0212(1)$ \\
10 & $-0.435947(3)$ & $0.00400(2)$ & $0.0211(1)$ \\
12 & $-0.435813(2)$ & $0.00411(2)$ & $0.0216(1)$ \\
14 & $-0.435749(2)$ & $0.00390(2)$ & $0.0205(1)$ \\
$16^{*}$ & $-0.435210(2)$ & $0.00593(2)$ & $0.0313(1)$ \\
    \end{tabular}
    \end{ruledtabular}
\end{table}
%%%%%%%%%%%%%%%%%%%%%%%%%%%%%%
%%%%%%%%%%%%%%%%%%%%%%%%%%%%%%
\begin{table}[]
\caption{Variational energy per site, variance per site and V-score ($V_{\text{odd}}$) of the ViT states in the ($\Gamma,S^z=0, P=-1$) sector after further optimization steps in the $A_2$ ($E_2$ for $L=4$) irrep of the $C_{6v}$ point group.}
    \label{tab:odd_and_gap}
    \begin{ruledtabular}
    \begin{tabular}{cccc}
$L$ & Energy per site & Variance per site & $V_{\text{odd}}$\\
\hline
4 & $-0.433729(5)$ & $0.00303(2)$ & $0.0161(1)$ \\
6 & $-0.435143(5)$ & $0.00229(3)$ & $0.0121(1)$ \\
8 & $-0.435677(3)$ & $0.00311(1)$ & $0.0164(1)$ \\
10 & $-0.435683(3)$ & $0.00386(2)$ & $0.0203(1)$ \\
12 & $-0.435704(3)$ & $0.00426(2)$ & $0.0224(1)$ \\
    \end{tabular}
    \end{ruledtabular}
\end{table}
%%%%%%%%%%%%%%%%%%%%%%%%%%%%%%

%%%%%%%%%%%%%%%%%%%%%%%
\begin{figure}[t]
    \centering
    \includegraphics[width=\columnwidth]{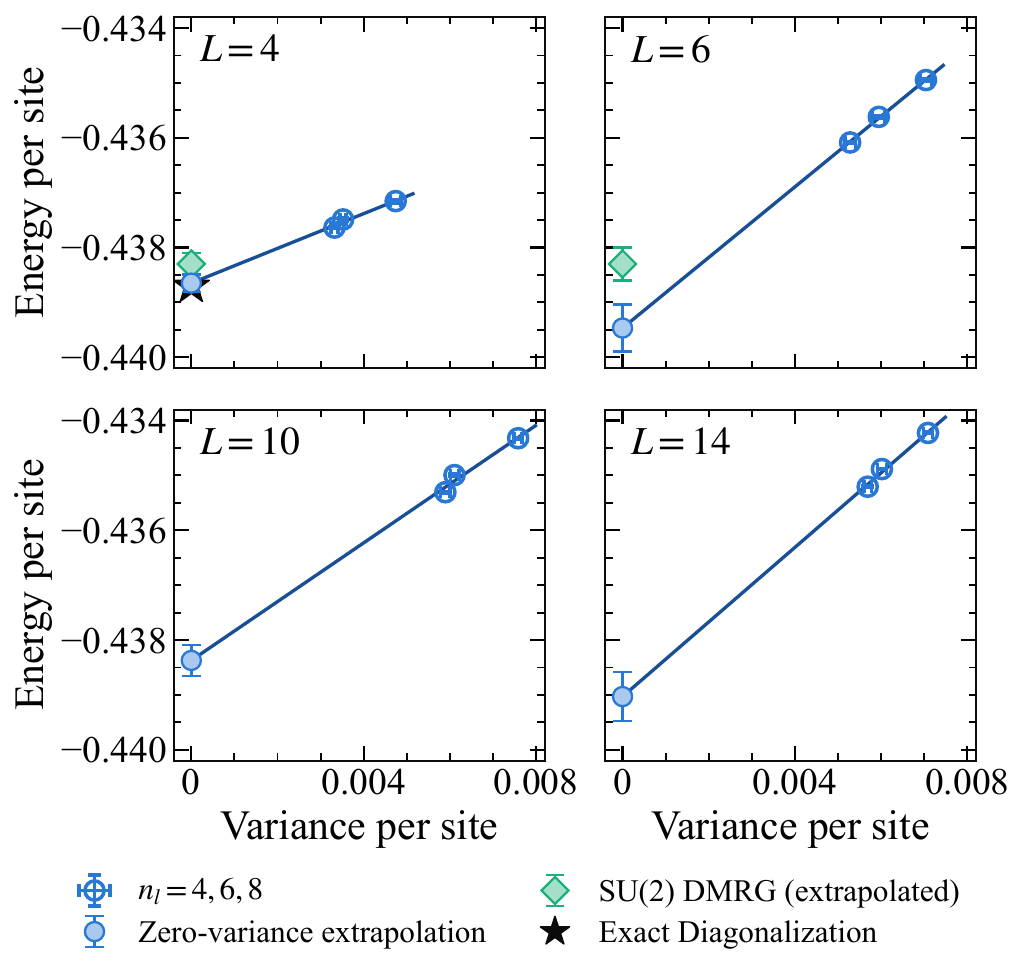}
    \caption{Energy per site against the variance per site of the even-parity ($P=+1$) states for $L=4$, $6$, $10$ and $14$, obtained with transformers with $n_l=4$, $6$ and $8$ layers before the point-group projection (open blue circles). Solid lines are linear fits, extrapolated to zero variance (filled blue circles). The error bars of the extrapolated energies are the standard deviation of the intercept over replicas of the data, in which energies and variances are displaced by Gaussian noise with the width of their error bars. The star marks the exact ground-state energy for $L=4$~\cite{lauchli2019exact}, and the green diamonds the extrapolated $SU(2)$ DMRG energies of the $L=4$ and $L=6$ tori~\cite{depenbrock2012dmrg}.
}
    \label{fig:varextrap}
\end{figure}
%%%%%%%%%%%%%%%%%%%%%%%

\paragraph*{Variance Extrapolations.}
In~\Cref{fig:varextrap}, we show how an extrapolation in the energy variance improves upon the variational energies of~\cref{tab:energy}.
For each system size we consider transformers with $n_l=4$, $6$ and $8$ layers (before the point-group projection) and plot their energy per site against the variance per site.
A linear fit is extrapolated to zero variance, and the uncertainty of the extrapolated energy is estimated by repeating the fit on Gaussian-distributed replicas of the data within their error bars.
For $L=4$ we obtain $-0.43865(16)$, consistent with the exact value $-0.438704$, whereas the  $SU(2)$ DMRG energy reported in Ref.~\cite{depenbrock2012dmrg} for the same torus, $-0.4383(2)$, lies two error bars above it.
For $L=6$ the extrapolation gives $-0.43947(43)$, while the $SU(2)$ DMRG energy of the same torus is $-0.4383(3)$~\cite{depenbrock2012dmrg}.
The extrapolated energies for $L=10$ and $14$ are $-0.43837(29)$, and $-0.43903(44)$, respectively.\\

%%%%%%%%%%%%%%%%%%%%%%%%%
\begin{figure}[t]
    \centering
    \includegraphics[width=\columnwidth]{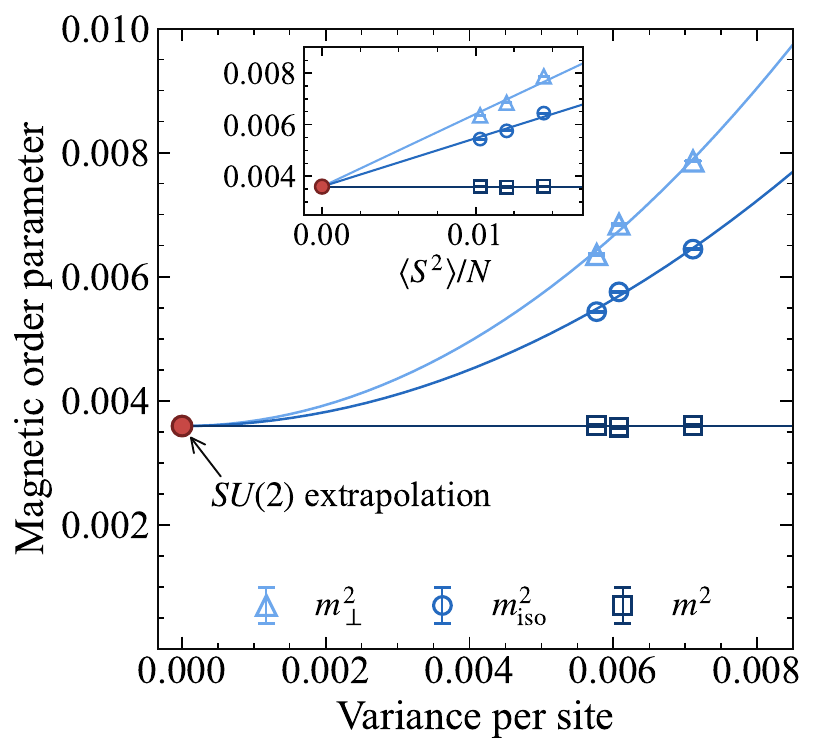}
    \caption{Magnetic order parameter computed from the $zz$ ($m^2$), isotropic ($m^2_{\text{iso}}$) and transverse ($m^2_{\perp}$) spin-spin correlations, as a function of the energy variance per site for $L=12$.
    The three data points of each estimator correspond to transformer wave functions with $n_l=4$, $6$ and $8$ layers; solid lines are fits to the data.
    Inset: the same data as a function of $\langle S^2 \rangle/N$. The zero-variance extrapolated value is reported as a red circle.}
    \label{fig:m2_SU2}
\end{figure}
%%%%%%%%%%%%%%%%%%%%%%%%%

\paragraph*{$SU(2)$ extrapolations of the structure factor.}
Here, we detail the calculation of the spin structure factor, providing evidence that the $zz$ component yields the best approximation to the full isotropic value.
We compare three estimators of the square magnetic order parameter, which differ in the spin components entering the peak of the structure factor:
the isotropic estimator $m^2_{\text{iso}}$, obtained from $S_{\text{iso}}(\boldsymbol{q})=\frac{1}{N}\sum_{i,j}e^{\mathrm{i}\boldsymbol{q}\cdot(\boldsymbol{r}_i-\boldsymbol{r}_j)}\langle\hat{\boldsymbol{S}}_i\cdot\hat{\boldsymbol{S}}_j\rangle$;
the transverse estimator $m^2_{\perp}$, obtained from $S_{\perp}(\boldsymbol{q})=\frac{3}{2N}\sum_{i,j}e^{\mathrm{i}\boldsymbol{q}\cdot(\boldsymbol{r}_i-\boldsymbol{r}_j)}\langle\hat{S}^x_i\hat{S}^x_j+\hat{S}^y_i\hat{S}^y_j\rangle$;
and the longitudinal estimator $m^2$ used in the main text, obtained from the $zz$ structure factor $S(\boldsymbol{q})=\frac{3}{N}\sum_{i,j}e^{\mathrm{i}\boldsymbol{q}\cdot(\boldsymbol{r}_i-\boldsymbol{r}_j)}\langle\hat{S}^z_i\hat{S}^z_j\rangle$.
The constant prefactors are chosen such that the three estimators coincide for an exactly $SU(2)$-symmetric state.
\Cref{fig:m2_SU2} shows the three estimators for three variational states of increasing accuracy, obtained with transformers of different depths, as a function of the energy variance per site, which we take as a measure of variational accuracy.
For each estimator, we fit the data and extrapolate to zero variance.
At the finite accuracy reached here, $m^2$ is already converged with respect to the variance, whereas $m^2_{\perp}$ and $m^2_{\text{iso}}$ are overestimated and decrease systematically toward it as the variance is reduced.
Accordingly, the extrapolations to zero variance, where $SU(2)$ symmetry is restored, yield values consistent with $m^2$ and well below the finite-variance estimates of $m^2_{\perp}$ and $m^2_{\text{iso}}$.
The inset shows that the deviation of $m^2_{\perp}$ and $m^2_{\text{iso}}$ from $m^2$ decreases together with $\langle \hat{S}^2\rangle/N$, confirming that it originates from the residual $SU(2)$-symmetry breaking of the variational state and is progressively suppressed as the variational accuracy improves.
These results justify using $m^2$ as our estimator of the magnetic order parameter.

%%%%%%%%%%%%%%%%%
% Supplemental Material
%%%%%%%%%%%%%%%%%
\clearpage
\onecolumngrid

\begin{center}
\textbf{\large Supplemental Material for\\[2pt]
``Gapless Nature of Spin Liquid in the $S=1/2$ Kagome Heisenberg Antiferromagnet with Neural-Network Quantum States''}
\end{center}

\setcounter{equation}{0}
\setcounter{figure}{0}
\setcounter{table}{0}
\renewcommand{\theequation}{S\arabic{equation}}
\renewcommand{\thefigure}{S\arabic{figure}}
\renewcommand{\thetable}{S\arabic{table}}
\renewcommand{\theHequation}{S\arabic{equation}}
\renewcommand{\theHfigure}{S\arabic{figure}}
\renewcommand{\theHtable}{S\arabic{table}}

\section*{Optimization Details}
To target the lowest singlet and the lowest triplet within the zero-momentum sector, we project the wave function onto even and odd spin-inversion parity, respectively, using the symmetrized variational states $\Psi^\pm_{\theta}(\boldsymbol{\sigma})=\Psi^{\mathrm{TR}}_\theta(\boldsymbol{\sigma})\pm\Psi^{\mathrm{TR}}_\theta(-\boldsymbol{\sigma})$, where $-\boldsymbol{\sigma}$ denotes the configuration with all spins flipped.
The variational parameters $\theta$ are optimized within the variational Monte Carlo (VMC) framework~\cite{becca2017quantum} by minimizing the variational energy $E^{\pm}_\theta=\langle\Psi^{\pm}_{\theta}|\hat H|\Psi^{\pm}_{\theta}\rangle/\langle\Psi^{\pm}_{\theta}|\Psi^{\pm}_{\theta}\rangle$, which is estimated via Markov chain Monte Carlo from configurations sampled according to the Born distribution $P^{\pm}_{\theta}(\boldsymbol{\sigma}) \propto |\Psi^{\pm}_{\theta}(\boldsymbol{\sigma})|^2$.
All optimizations are performed in the $S^z=0$ sector, which is enforced by the transition rule of the Monte Carlo sampling.
The energy is minimized with the MARCH optimizer~\cite{gu2026solving}, which builds on SR~\cite{sorella1998green,sorella2005wave} in the formulation suited to a number of variational parameters much larger than the number of samples~\cite{chen2024minsr,rende2024simple,goldshlager2024spring}.
We use MARCH parameters $\mu=0.95$ and $\beta=0.995$, a diagonal shift of $10^{-4}$, and $M=2^{14}$ Monte Carlo samples.
Unless stated otherwise, all optimizations use a transformer with $n_l=8$ layers, $h=12$ attention heads, and embedding dimension $d=72$.

First, we perform an initial optimization of $10^4$ steps with a learning rate decaying from $5\times10^{-3}$ to $5\times10^{-4}$ following a cosine schedule~\cite{loschilov2017cosine}.
We then restore the $C_{6v}$ point-group symmetry by projection.
In the following, we denote with $|\Psi_P\rangle$ the optimized spin-inversion-symmetrized state of either even or odd parity.
To select the irreducible representation (irrep) $\lambda$, we compute the weights $w_\lambda=\langle\Psi_P|\hat P_\lambda|\Psi_P\rangle/\langle\Psi_P|\Psi_P\rangle$ in all irreps of $C_{6v}$, where $\hat P_\lambda$ is the projector onto $\lambda$ and $\sum_\lambda w_\lambda=1$, and project onto the irrep with the largest weight.
For all even states, the largest weight is in $A_1$, with $w_{A_1}$ close to one, whereas for the odd states it is in $A_2$ for $L\geq6$ and in $E_2$ for $L=4$.
The point-group-projected state in irrep $\lambda$ is $\Psi_\lambda(\boldsymbol{\sigma})=\sum_{g\in C_{6v}}\chi_\lambda(g)\,\Psi_P(g\boldsymbol{\sigma})$, where $\chi_\lambda(g)$ is the character of $g$ in $\lambda$ and $g\boldsymbol{\sigma}$ denotes the configuration transformed by $g$.
Starting from the parameters obtained in the initial optimization, we optimize $|\Psi_\lambda\rangle$ for a further $300$ steps at a fixed learning rate of $3\times10^{-3}$, with all other settings unchanged.

\end{document}